\UseRawInputEncoding

\documentclass[10pt, prb, aps, twocolumn, showpacs, citeautoscript, floatfix, reprint, amsmath, amssymb, notitlepage, superscriptaddress]{revtex4-1}

\usepackage{graphicx}
\usepackage{stmaryrd}
\usepackage{rotating}
\usepackage{amsmath}
\usepackage{amsfonts}
\usepackage{amssymb}
\usepackage{wasysym}
\usepackage[countmax]{subfloat}
\usepackage{dcolumn} 
\usepackage{bm} 
\usepackage{color}

\usepackage{float}
\usepackage{latexsym}

\begin{document}

\title{Berry curvature and quantum metric in copper-substituted lead phosphate apatite}



\author{Wei Chen}

\affiliation{Department of Physics, PUC-Rio, 22451-900 Rio de Janeiro, Brazil}

\date{\rm\today}

\begin{abstract}

The recent discovery of copper-substituted lead phosphate apatite, also known as LK-99, has caught much attention owing to certain experimental evidence of room-temperature superconductivity, although this claim is currently under intensive debate. Be it superconducting or not, we show that the normal state of this material has peculiar quantum geometrical properties that may be related to the magnetism and the mechanism for flat band superconductivity. Based on a recently proposed spinless two-band tight-binding model for the Pb-Cu hexagonal lattice subset of the crystalline structure, which qualitatively captures the two flat bands in the band structure, we elaborate the highly anisotropic Berry curvature and quantum metric in the regions of Brillouin zone where one flat band is above and the other below the Fermi surface. In these regions, the Berry curvature has a pattern in the planar momentum that remains unchanged along the out-of-plane momentum. Moreover, the net orbital magnetization contributed from the Berry curvature is zero, signifying that the magnetism in this material should come from other sources. The quantum metric has a similar momentum dependence, and its two planar components are found to be roughly the same but the out-of-plane component vanishes, hinting that the superfluid stiffness of the flat band superconductivity, shall it occur, may be quite anisotropic.

\end{abstract}

\maketitle

\section{Introduction}

The copper-substituted lead phosphate apatite CuPb$_{9}$(PO$_{4}$)$_{6}$OH$_{2}$ (also known as LK-99) discovered recently has stirred a fair amount of interest even in the general public, owing to the claim of possible observation of room-temperature superconductivity\cite{Lee23_1,Lee23_2}. This claim is based on several preliminary experimental results that are yet to be verified, including zero resistance transport and magnetic levitation. Despite these appealing initial results, there is a growing concern that the observed magnetic levitation may actually be attributed to diamagnetism, and therefore the magnetic properties of this material seems to be an equally important issue. Meanwhile, on the theoretical side, besides several attempts to formulate a pairing mechanism for the superconductivity, the density-functional theory (DFT) calculations have greatly helped to understand the normal state band structure of this material\cite{Griffin23,Lai23,Si23,Kurleto23,CabezasEscares23}. The important feature revealed by the DFT calculation is a pair of flat bands around the Fermi surface with a bandwidth as narrow as $\sim 0.1$eV in all momentum directions, which presumably have very peculiar electronic properties. Based on the DFT results, Tavakol and Scaffidi further propose a tight-binding model solely describing the Cu $d_{zx}$ and $d_{zy}$ orbitals of the Pb-Cu subset of the crystal structure that forms a hexagonal lattice, which qualitatively captures the flat bands and dramatically simplifies the problem\cite{Tavakol23}.

In this paper, we elaborate the nontrivial quantum geometrical properties of the aforementioned two-band tight-binding model, which may help to shed some light on the magnetism and mechanisms for superconductivity. Focusing entirely on the metallic normal state of the material, we point out that there are regions in the Brillouin zone (BZ) where the band structure has one flat band below and the other above the Fermi energy. In addition, the tight-binding model takes the form of a Dirac Hamiltonian, and therefore in these regions the Hamiltonian bears a striking similarity with those of the topological insulators (TIs) and topological superconductors (TSCs)\cite{Schnyder08,Ryu10,Chiu16,Hasan10,Qi11}. Consequently, the Berry curvature and quantum metric that are well investigated in TIs and TSCs, both are measures of the variation of the valence band states in momentum space, can be nonzero in these regions. In particular, Berry curvature is known to contribute to an orbital magnetization\cite{Chang96,Xiao10}, which serves as a strong motivation to investigate the momentum profile of the Berry curvature within this tight-binding model.

In contrast, our motivation for investigating the quantum metric comes from its relation with the flat band superconductivity. In another equally controversial superconducting material discovered not long ago, namely the twisted bilayer graphene (TBLG)\cite{Cao18}, the misalignment of the two graphene layers also produce flat bands in a small fraction of the BZ that are presumably responsible for the superconductivity\cite{Cao18_2}. A class of theory suggests that the superfluid stiffness of the superconducting order parameter in TBLG comes from the normal state quantum metric of the flat band\cite{Peotta15,Julku16,Liang17,HerzogArbeitman22,Torma22}, which serves as our motivation to investigate the quantum metric. It should be noted, though, that TBLG has much flatter bands $\sim 0.01$meV than LK-99 $\sim 0.1$eV, although the later extends through the entire BZ. Our calculation reveals that the quantum metric LK-99 along the out-of-plane direction vanishes, hinting that the flat band superconductivity, shall it manifest, may be very anisotropic. 


The structure of the paper is organized in the following manner. In Sec.\ref{sec:tight_binding}, we review the two-band tight-binding model, and in Sec.~\ref{sec:Dirac_model_general_formalism} we mention the formula for the Berry curvature, orbital magnetization, and quantum metric in a generic $2\times 2$ Dirac Hamiltonian. Analytical results around several momentum points and the full profile of these geometrical quantities are shown in Sec.~\ref{sec:analytical_numerical_results}. Finally, in Sec.~\ref{sec:conclusions}, we summarize our results and discuss their implications.

\section{Berry curvature and quantum metric in the tight-binding model \label{sec:Berry_metric}}

\subsection{Spinless tight-binding model for the Pb-Cu hexagonal lattice \label{sec:tight_binding}}

Our starting point is the $2\times 2$ spinless tight-binding model that takes the form of a Dirac Hamiltonian\cite{Tavakol23} 
\begin{eqnarray}
H_{0}({\bf k})=(d_{0}-\mu){\bf I}+{\bf d}\cdot{\boldsymbol\sigma},
\end{eqnarray}
where $\left\{\sigma_{1},\sigma_{2},\sigma_{3}\right\}$ are Pauli matrices. The ${\bf d}$-vector is parametrized by 
\begin{eqnarray}
&&d_{0}({\bf k})=-\frac{1}{4}\sum_{i=1}^{3}\cos({\bf k\cdot b}_{i})-\frac{3}{2}-2t_{z}\cos k_{z},
\nonumber \\
&&d_{1}({\bf k})=-\frac{\sqrt{3}}{4}\left[-\cos({\bf k\cdot b}_{1})+\cos({\bf k\cdot b}_{2})\right],
\nonumber \\
&&d_{2}({\bf k})=-\frac{\sqrt{3}}{4}\sum_{i=1}^{3}\sin({\bf k\cdot b}_{i}),
\nonumber \\
&&d_{3}({\bf k})=-\frac{1}{2}\cos({\bf k\cdot b}_{3})+\frac{1}{4}\left[\cos({\bf k\cdot b}_{1})+\cos({\bf k\cdot b}_{2})\right].
\nonumber \\
\end{eqnarray}
In units of bond length $a$ of the honeycomb lattice, the 2D vectors in real space are parametrized by 
\begin{eqnarray}
{\bf b}_{1}=\left(-\frac{\sqrt 3}{2},-\frac{3}{2}\right),\;\;
{\bf b}_{2}=\left(-\frac{\sqrt 3}{2},\frac{3}{2}\right),\;\;
{\bf b}_{3}=\left(\sqrt{3},0\right).
\nonumber \\
\end{eqnarray}
Note that the Hamiltonian is time-reversal (TR) symmetric $TH_{0}({\bf k})T^{-1}=H_{0}(-{\bf k})$, with the TR operator implemented by complex conjugation. This seems to already imply that the band structure alone cannot produce a magnetization. Secondly, we also note that the flat bands are entirely contributed from the hopping between Cu substitutions. This is very different from another well-known mechanism for the flat bands where the substituted atoms or periodic vacancies produce unequal amount of sublattices on a bipartite lattice\cite{Sutherland86,Lieb89}, which has also been propose to stabilize superconductivity\cite{deSousa23_flat_band}.


\subsection{Berry curvature, orbital magnetization, and quantum metric \label{sec:Dirac_model_general_formalism}}

The Dirac form of the Hamiltonian and the fact that $(d_{1},d_{2},d_{3})$ only depend on the planar momentum $(k_{x},k_{y})$ motivate us to investigate the Berry curvature and quantum metric of this model, as it bears a striking similarity with the 2D Chern insulator\cite{Hasan10,Qi11}. However, an obvious difference is that the two-band model is metallic\cite{Tavakol23}, in contrast to the Chern insulator that has a band gap throughout the BZ. As a result, the Berry curvature and quantum metric of this two-band model is only defined in the regions of BZ where the lower band $E_{-}({\bf k})$ is below the Fermi surface and hence serves as a valence band, and the higher band $E_{+}({\bf k})$ is above the Fermi surface and hence plays the role of a conduction band (in the regions where the two bands are both above or both below the Fermi surface, the Berry curvature and quantum metric are zero within this two-band model). In these gapped regions, we denote the valence and conduction band states to be $|n\rangle$ and $|m\rangle$, respectively. Denoting the derivative over momentum as $\partial_{\mu}=\partial/\partial k_{\mu}$ and $d\equiv\sqrt{d_{1}^{2}+d_{2}^{2}+d_{3}^{2}}$, our interest is the Berry curvature of the filled valence band state $|n\rangle$, which is calculated by 
\begin{eqnarray}
&&\Omega_{\mu\nu}=i\langle\partial_{\mu}n|\partial_{\nu}n\rangle
-i\langle\partial_{\nu}n|\partial_{\mu}n\rangle
\nonumber \\
&&=\frac{1}{2d^{3}}\varepsilon^{ijk}d_{i}\partial_{\mu}d_{j}\partial_{\nu}d_{k},
\end{eqnarray}
where $\varepsilon^{ijk}$ is the fully antisymmetric Levi-Civita symbol. As we shall see below, the Berry curvature has only the $\Omega_{xy}$ component, so it contributes to an orbital magnetization along ${\hat{\bf z}}$ direction at momentum ${\bf k}$ described by\cite{Chang96,Xiao10}
\begin{eqnarray}
&&m^{z}({\bf k})=-i\frac{e}{2\hbar}\left[\langle\partial_{x}n|(H-E_{-})|\partial_{y}n\rangle\right.
\nonumber \\
&&\left.-\langle\partial_{y}n|(H-E_{-})|\partial_{x}n\rangle\right]
\nonumber \\
&&=\frac{e}{2\hbar}(E_{+}-E_{-})\Omega_{xy}
\nonumber \\
&&=-\left(\frac{e}{\hbar}\right)\frac{1}{2d^{2}}\varepsilon^{ijk}d_{i}\partial_{\mu}d_{j}\partial_{\nu}d_{k}.
\end{eqnarray}
The net orbital magnetization of the material is then given by the momentum integration of $m^{z}({\bf k})$.

On the other hand, the quantum metric is defined from the overlap of the valence band state at slightly different momenta\cite{Provost80} $|\langle n({\bf k})|n({\bf k+\delta k})\rangle|=1-g_{\mu\nu}\delta k_{\mu}\delta k_{\nu}$, which in the $2\times 2$ Dirac Hamiltonian has the expression\cite{vonGersdorff21_metric_curvature,Chen23_dressed_Berry_metric,deSousa23_fidelity_marker}
\begin{eqnarray}
&&g_{\mu\nu}=\frac{1}{2}\langle\partial_{\mu}n|\partial_{\nu}n\rangle
+\frac{1}{2}\langle\partial_{\nu}n|\partial_{\mu}n\rangle
-\langle\partial_{\mu}n|m\rangle\langle m|\partial_{\nu}n\rangle
\nonumber \\
&&=\frac{1}{4d^{2}}\left\{\sum_{i=1}^{3}\partial_{\mu}d_{i}\partial_{\nu}d_{i}
-\partial_{\mu}d\,\partial_{\nu}d\right\}.
\end{eqnarray}
Note that the $d_{0}$ component does not influence the eigenstates, so the Berry curvature and quantum metric are entirely determined by $\left\{d_{1},d_{2},d_{3}\right\}$. However, the $d_{0}$ component does influence which states are above and below the Fermi surface and hence whether Berry curvature and quantum metric are nonzero, as we shall see below.

\subsection{Analytical and numerical results \label{sec:analytical_numerical_results}}

Analytical results near high-symmetry points (HSPs) of the BZ can be given to draw some theoretical understanding. In particular, the $K$ and $K'$ points defined by (in units of $\hbar/a$)
\begin{eqnarray}
{\bf K}=\left(\frac{2\pi}{3\sqrt{3}},\frac{2\pi}{3}\right),\;\;\;{\bf K}'=\left(-\frac{2\pi}{3\sqrt{3}},\frac{2\pi}{3}\right),
\end{eqnarray}
on the hexagonal plane of $k_{z}=0$ belong to the gapped regions, and are of particular interest since they are analogous to those in graphene (even though graphene is not gapped), so we seek for analytical result near these points. A straightforward expansion of the momentum near these points ${\bf k}={\bf K}^{(\prime)}+\delta{\bf k}$ yields
\begin{eqnarray}
&&K\;{\rm point}:\;\;\;d_{1}=-\frac{9}{8}\delta k_{y},\;\;\;
d_{2}=-\frac{9}{8},\;\;\;
d_{3}=\frac{9}{8}\delta k_{x},
\nonumber \\
&&K'\;{\rm point}:\;\;\;d_{1}=\frac{9}{8}\delta k_{y},\;\;\;
d_{2}=\frac{9}{8},\;\;\;
d_{3}=-\frac{9}{8}\delta k_{x}.
\end{eqnarray}
Firstly, converting $\delta k_{x}=\delta k\sin\theta\cos\phi$ and $\delta k_{y}=\delta k\sin\theta\sin\phi$ into spherical coordinate, the vanishing integration of the Berry curvature over a spherical surface of radius $\delta k$
\begin{eqnarray}
\frac{1}{2\pi}\int_{0}^{2\pi}d\phi\int_{0}^{\pi}d\theta\,\Omega_{\theta\phi}=0,
\end{eqnarray}
indicating that there is no topological charge at $K$ and $K'$ points, defined analogous to that of the 3D Weyl semimetals\cite{Armitage18,Yan17,Lv21}. On the other hand, in the Cartesian coordinate $\partial_{\mu}=\partial/\partial\delta k_{\mu}$, the Berry curvature and quantum metric are
\begin{eqnarray}
&&\Omega_{xy}=\pm\frac{1}{2(1+\delta k^{2})^{3/2}},
\nonumber \\
&&g_{\mu\nu}=\left(\begin{array}{cc}
g_{xx} & g_{xy} \\
g_{yx} & g_{yy}
\end{array}\right)
\nonumber \\
&&=
\frac{1}{4(1+\delta k^{2})^{2}}\left(\begin{array}{cc}
1+\delta k_{y}^{2} & -\delta k_{x}\delta k_{y} \\
-\delta k_{x}\delta k_{y} & 1+\delta k_{x}^{2}
\end{array}\right),
\label{KKP_Berry_metric}
\end{eqnarray}
where the upper and lower signs in $\Omega_{xy}$ are for the $K$ and $K'$ points, respectively. The results in Eq.~(\ref{KKP_Berry_metric}) satisfies a metric-curvature correspondence 
\begin{eqnarray}
\sqrt{\det g_{\mu\nu}}=\frac{1}{2}|\Omega_{xy}|=\frac{1}{4(1+\delta k^{2})^{3/2}},
\end{eqnarray}
a relation that is found to be valid for TIs and TSCs described by Dirac models in any dimension and symmetry class\cite{vonGersdorff21_metric_curvature}. It should be noted, though, that the material at hand is 3D but the relation is satisfied in a 2D manner, in the sense that $g_{\mu\nu}$ has no $g_{zz}$ component and the determinant is completely calculated from the planar components.

\begin{figure*}[ht]
\begin{center}
\includegraphics[clip=true,width=1.8\columnwidth]{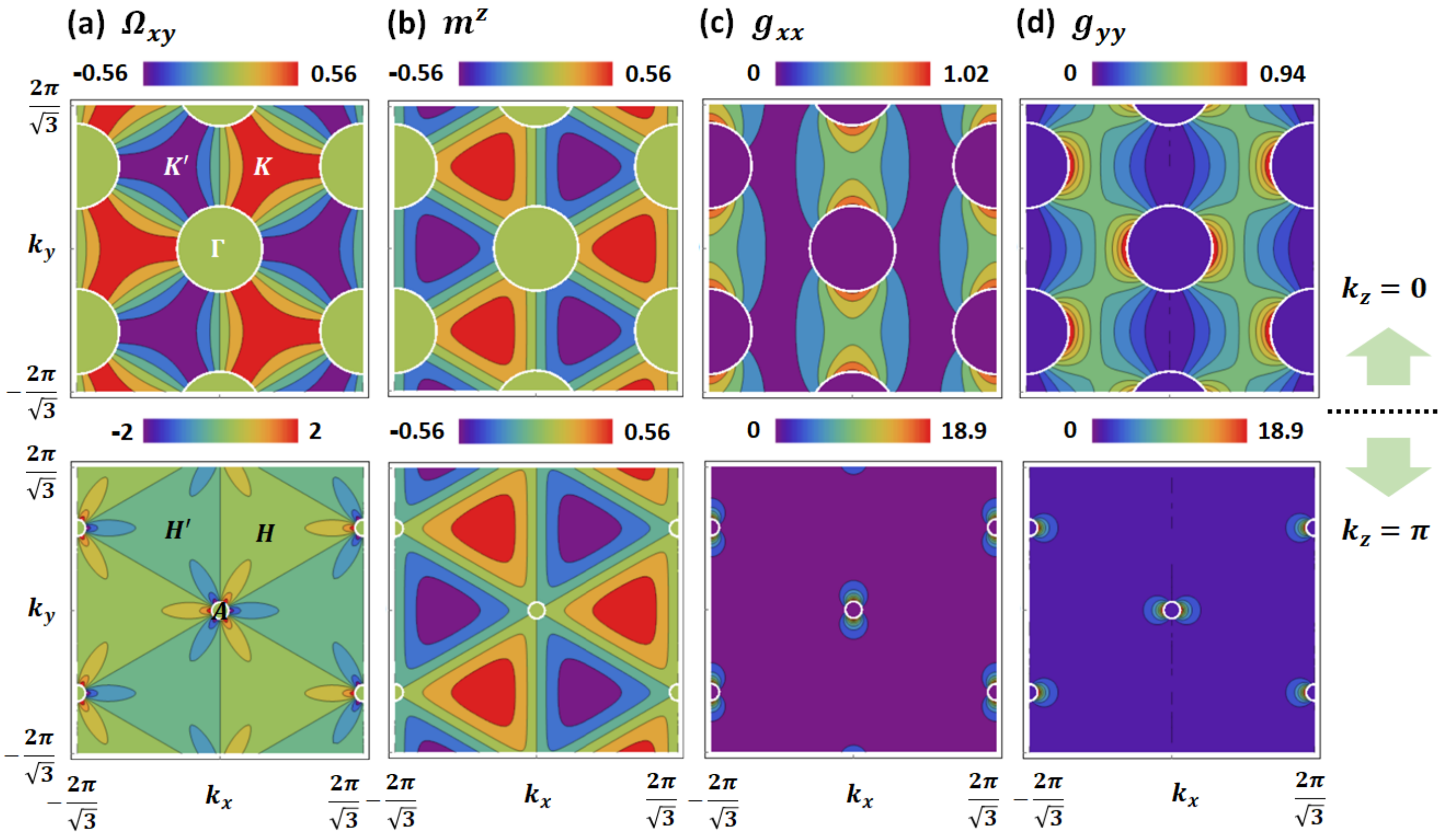}
\caption{(a) Berry curvature $\Omega_{xy}$, (b) orbital magnetization $m^{z}$, (c) quantum metric component $g_{xx}$, and (d) component $g_{yy}$ in the momentum space of the two-band tight-binding model plotted as a function of $(k_{x},k_{y})$ for $k_{z}=0$ (top panels) and $k_{z}=\pi$ (bottom panels). The values of these quantities are zero at the area around the $\Gamma$ point where both flat bands are below the Fermi surface, and this area shrinks when $k_{z}$ changes from $0$ to $\pi$. Note that the scale of plotting in each panel is different, as indicated by the color bar. The HSPs $\left\{\Gamma,K,K',A,H,H'\right\}$ are labeled in panel (a). } 
\label{fig:Berry_metric_results}
\end{center}
\end{figure*}

Numerical results for $\Omega_{xy}$, $m^{z}$, $g_{xx}$ and $g_{yy}$ are shown in Fig.~\ref{fig:Berry_metric_results}, where we plot them as functions of planar momentum $(k_{x},k_{y})$ at two different values of out-of-plane momentum $k_{z}=0$ and $\pi$. Notice that these quantities should not evolve with $k_{z}$ since the eigenstates $|n\rangle$ and $|m\rangle$ are determined by $(d_{1},d_{2},d_{3})$ that do not depend on $k_{z}$. Nevertheless, as mentioned above, these quantities are nonzero only when $E_{-}<0$ and $E_{+}>0$, so they are zero in the region surrounding the $\Gamma-A$ line  where $\left\{E_{+},E_{-}\right\}<0$. This region shrinks as $k_{z}$ moves from $0$ to $\pi$, which also chages the overall scale of the plots between the upper panels and the lower panels in Fig.~\ref{fig:Berry_metric_results}, since these quantites have a larger magnitude around the $\Gamma-A$ line that is gradually revealed as $k_{z}$ moves from $0$ to $\pi$ (moving from $\Gamma$ to $A$). At the boundary of this $\left\{E_{+},E_{-}\right\}<0$ region, these quantites has a discrete jump from zero to a finite value, which many have some interesting consequences that shall be further explored.

Concerning the momentum dependence of the Berry curvature $\Omega_{xy}$ and the orbital magnetization $m^{z}$ shown in Fig.~\ref{fig:Berry_metric_results} (a) and (b), we find that their values around $K-H$ and $K'-H'$ lines are always of equal magnitude and opposite signs, and hence cancel out after a momentum integration, indicating no net orbital moment. This behavior is very similar to that in transition metal dichalcogenides (TMDs), which have a two-dimensional (2D) hexagonal BZ and also show this type of cancellation of orbital moment and Berry curvature\cite{Feng12,Liu13}. In addition, the Berry curvature around the $\Gamma-A$ line is actually much higher than at $K-H$ and $K'-H'$ lines, and maintains the pattern of alternating signs and lobes of the six-fold symmetry of the BZ, although they are only revealed as $k_{z}$ approaches $\pi$ such that the $E_{-}<0$ and $E_{+}>0$ condition is satisfied (as moving from $\Gamma$ to $A$), as discussed above.

On the other hand, the quantum metric $g_{xx}$ and $g_{yy}$ shown in Fig.~\ref{fig:Berry_metric_results} (c) and (d) are both positively defined, and their values are the same at $K-H$ and $K'-H'$ lines. The quantities are also found to be much more enhanced around the $\Gamma-A$ line, with a pattern of $g_{xx}$ that has lobes along $\pm k_{y}$ directions and those of $g_{yy}$ are along $\pm k_{x}$ directions, and the lobes of $g_{xx}$ and $g_{yy}$ are roughly of the same magnitude. To further quantify the difference between $g_{xx}$ and $g_{yy}$, we perform a momentum-integration of these quantities over the 3D BZ to obtain 
\begin{eqnarray}
{\cal G}_{\mu\mu}=\int\frac{d^{3}{\bf k}}{\tilde{V}_{BZ}}\,g_{\mu\mu},
\end{eqnarray}
which is a measure of the average distance between valence band states $|n\rangle$ along the $k_{\mu}$ direction in the $E_{-}<0$ and $E_{+}>0$ regions of the 3D BZ, a quatity that has been called the fidelity number\cite{deSousa23_fidelity_marker}. Here $\tilde{V}=16\pi^{3}/3\sqrt{3}$ is the dimensionless volume of the 3D BZ. The numerical result yields ${\cal G}_{xx}\approx 0.3742$ and ${\cal G}_{yy}\approx 0.3734$ in units of $\hbar/c$, where $c$ is the lattice constant along ${\hat{\bf z}}$ direction, indicating that the two planar directions have roughly the same quantum geometrical properties. In contrast, quantum metric along the ${\hat{\bf z}}$ direction vanishes everywhere $g_{zz}=g_{xz}=g_{yz}=0$, and hence if the quantum metric really plays a role in some pairing mechanisms, such as the flat band superconductivity\cite{Peotta15,Julku16,Liang17,HerzogArbeitman22,Torma22}, then the superconductivity likely only occurs in the ${\hat{\bf x}}$ and ${\hat{\bf y}}$ directions despite the band structure is flat in all three directions.

Finally, we remark that in semiconductors and insulators, one can define a Berry curvature spectral function\cite{Chen23_dressed_Berry_metric} $\Omega_{xy}({\bf k},\omega)$ that frequency-integrates to the Berry curvature $\Omega_{xy}({\bf k})=\int d\omega\,\Omega_{xy}({\bf k},\omega)$ and momentum-integrates to the Chern number spectral function ${\cal C}(\omega)=\int\frac{d^{2}{\bf k}}{(2\pi)^{2}}\Omega_{xy}({\bf k},\omega)$, with the later being related to the circular dichroism of 2D materials\cite{Molignini22_Chern_marker}. Likewisely, one can also introduce a quantum metric spectral function\cite{Chen23_dressed_Berry_metric} $g_{\mu\nu}({\bf k},\omega)$ that frequency-integrates to the quantum metric $g_{\mu\mu}({\bf k})=\int d\omega\,g_{\mu\mu}({\bf k},\omega)$ and is proportional the optical conductivity $g_{\mu\mu}({\bf k},\omega)\propto\sigma_{\mu\mu}({\bf k},\omega)/\omega$ at momentum ${\bf k}$, which can possibly be detected in pump-probe type of experiments\cite{vonGersdorff21_metric_curvature}. Furthermore, one can integrate the quantum metric spectral function over momentum to obtain what we call the fidelity number spectral function ${\cal G}_{\mu\mu}(\omega)=\int\frac{d^{D}{\bf k}}{(2\pi)^{D}}g_{\mu\mu}({\bf k},\omega)\propto \sigma_{\mu\mu}(\omega)/\omega$ that corresponds to optical conductivity (and hence the optical absorption power) $\sigma_{\mu\mu}(\omega)$ measured in real space divided by frequency, which serves as a practical way to measure the quantum geometrical properties of insulators and semiconductors\cite{deSousa23_fidelity_marker}. These spectral functions, however, do not apply to the present work simply because the material under question is metallic, and hence does not have optical absorption since infrared light cannot penetrate through. It remains to be elucidated whether there is some simple methods to detect the quantum metric and Berry curvature for this material.

\section{Conclusions \label{sec:conclusions}}

In summary, we elaborate the intriguing momentum space quantum geometrical properties of the Pb-Cu hexagonal lattice tight-binding model for the normal state of LK-99, in an attempt to shed some light on the magnetism and the possibility of flat band superconductivity in this material. Our investigation focuses on the regions in the BZ where there is one flat band below $E_{-}<0$ and the other above $E_{+}>0$ the Fermi energy, in which many aspects in TIs and TSCs can be applied owing to the Dirac form of the Hamiltonian. Our calculation reveals that the Berry curvature of this tight-binding model only has the $\Omega_{xy}$ component and orbital magnetization only has the $m^{z}$ component, and both are found to inherit the six-fold symmetric pattern of the hexagonal lattice, with opposite signs between $K-H$ and $K'-H'$ lines and consequently momentum-integrates to zero, yielding no orbital magnetization and suggesting that the magnetism of this material should come from other sources. The quantum metric in the ${\hat{\bf x}}$ and ${\hat{\bf y}}$ directions roughly have the same magnitude, and are largely concentrated near the $\Gamma-A$ line. Moreover, the quantum metric along the ${\hat{\bf z}}$ direction vanishes, leading us to conjecture that shall the metric really play a role in some pairing mechanisms, such as the flat band superconductivity, then the resulting superconductivity may be highly anisotropic despite the bands are flat in all three directions. Further theoretical and experimental investigations are certainly needed to clarify the roles of these anisotropic Berry curvature and quantum metric, which await to be further explored.

\bibliography{Literatur}

\begin{thebibliography}{35}%
\makeatletter
\providecommand \@ifxundefined [1]{%
 \@ifx{#1\undefined}
}%
\providecommand \@ifnum [1]{%
 \ifnum #1\expandafter \@firstoftwo
 \else \expandafter \@secondoftwo
 \fi
}%
\providecommand \@ifx [1]{%
 \ifx #1\expandafter \@firstoftwo
 \else \expandafter \@secondoftwo
 \fi
}%
\providecommand \natexlab [1]{#1}%
\providecommand \enquote  [1]{``#1''}%
\providecommand \bibnamefont  [1]{#1}%
\providecommand \bibfnamefont [1]{#1}%
\providecommand \citenamefont [1]{#1}%
\providecommand \href@noop [0]{\@secondoftwo}%
\providecommand \href [0]{\begingroup \@sanitize@url \@href}%
\providecommand \@href[1]{\@@startlink{#1}\@@href}%
\providecommand \@@href[1]{\endgroup#1\@@endlink}%
\providecommand \@sanitize@url [0]{\catcode `\\12\catcode `\$12\catcode
  `\&12\catcode `\#12\catcode `\^12\catcode `\_12\catcode `\%12\relax}%
\providecommand \@@startlink[1]{}%
\providecommand \@@endlink[0]{}%
\providecommand \url  [0]{\begingroup\@sanitize@url \@url }%
\providecommand \@url [1]{\endgroup\@href {#1}{\urlprefix }}%
\providecommand \urlprefix  [0]{URL }%
\providecommand \Eprint [0]{\href }%
\providecommand \doibase [0]{http://dx.doi.org/}%
\providecommand \selectlanguage [0]{\@gobble}%
\providecommand \bibinfo  [0]{\@secondoftwo}%
\providecommand \bibfield  [0]{\@secondoftwo}%
\providecommand \translation [1]{[#1]}%
\providecommand \BibitemOpen [0]{}%
\providecommand \bibitemStop [0]{}%
\providecommand \bibitemNoStop [0]{.\EOS\space}%
\providecommand \EOS [0]{\spacefactor3000\relax}%
\providecommand \BibitemShut  [1]{\csname bibitem#1\endcsname}%
\let\auto@bib@innerbib\@empty
\bibitem [{\citenamefont {Lee}\ \emph {et~al.}(2023{\natexlab{a}})\citenamefont
  {Lee}, \citenamefont {Kim},\ and\ \citenamefont {Kwon}}]{Lee23_1}%
  \BibitemOpen
  \bibfield  {author} {\bibinfo {author} {\bibfnamefont {S.}~\bibnamefont
  {Lee}}, \bibinfo {author} {\bibfnamefont {J.-H.}\ \bibnamefont {Kim}}, \ and\
  \bibinfo {author} {\bibfnamefont {Y.-W.}\ \bibnamefont {Kwon}},\ }\href@noop
  {} {\bibfield  {journal} {\bibinfo  {journal} {arXiv:2307.12008}\ } (\bibinfo
  {year} {2023}{\natexlab{a}})}\BibitemShut {NoStop}%
\bibitem [{\citenamefont {Lee}\ \emph {et~al.}(2023{\natexlab{b}})\citenamefont
  {Lee}, \citenamefont {Kim}, \citenamefont {Kim}, \citenamefont {Im},
  \citenamefont {An},\ and\ \citenamefont {Auh}}]{Lee23_2}%
  \BibitemOpen
  \bibfield  {author} {\bibinfo {author} {\bibfnamefont {S.}~\bibnamefont
  {Lee}}, \bibinfo {author} {\bibfnamefont {J.}~\bibnamefont {Kim}}, \bibinfo
  {author} {\bibfnamefont {H.-T.}\ \bibnamefont {Kim}}, \bibinfo {author}
  {\bibfnamefont {S.}~\bibnamefont {Im}}, \bibinfo {author} {\bibfnamefont
  {S.}~\bibnamefont {An}}, \ and\ \bibinfo {author} {\bibfnamefont {K.~H.}\
  \bibnamefont {Auh}},\ }\href@noop {} {\bibfield  {journal} {\bibinfo
  {journal} {arXiv:2307.12037}\ } (\bibinfo {year}
  {2023}{\natexlab{b}})}\BibitemShut {NoStop}%
\bibitem [{\citenamefont {Griffin}(2023)}]{Griffin23}%
  \BibitemOpen
  \bibfield  {author} {\bibinfo {author} {\bibfnamefont {S.~M.}\ \bibnamefont
  {Griffin}},\ }\href@noop {} {\bibfield  {journal} {\bibinfo  {journal}
  {arXiv:2307.16892}\ } (\bibinfo {year} {2023})}\BibitemShut {NoStop}%
\bibitem [{\citenamefont {Lai}\ \emph {et~al.}(2023)\citenamefont {Lai},
  \citenamefont {Li}, \citenamefont {Liu}, \citenamefont {Sun},\ and\
  \citenamefont {Chen}}]{Lai23}%
  \BibitemOpen
  \bibfield  {author} {\bibinfo {author} {\bibfnamefont {J.}~\bibnamefont
  {Lai}}, \bibinfo {author} {\bibfnamefont {J.}~\bibnamefont {Li}}, \bibinfo
  {author} {\bibfnamefont {P.}~\bibnamefont {Liu}}, \bibinfo {author}
  {\bibfnamefont {Y.}~\bibnamefont {Sun}}, \ and\ \bibinfo {author}
  {\bibfnamefont {X.-Q.}\ \bibnamefont {Chen}},\ }\href@noop {} {\bibfield
  {journal} {\bibinfo  {journal} {arXiv:2307.16040}\ } (\bibinfo {year}
  {2023})}\BibitemShut {NoStop}%
\bibitem [{\citenamefont {Si}\ and\ \citenamefont {Held}(2023)}]{Si23}%
  \BibitemOpen
  \bibfield  {author} {\bibinfo {author} {\bibfnamefont {L.}~\bibnamefont
  {Si}}\ and\ \bibinfo {author} {\bibfnamefont {K.}~\bibnamefont {Held}},\
  }\href@noop {} {\bibfield  {journal} {\bibinfo  {journal} {arXiv:2308.00676}\
  } (\bibinfo {year} {2023})}\BibitemShut {NoStop}%
\bibitem [{\citenamefont {Kurleto}\ \emph {et~al.}(2023)\citenamefont
  {Kurleto}, \citenamefont {Lany}, \citenamefont {Pashov}, \citenamefont
  {Acharya}, \citenamefont {van Schilfgaarde},\ and\ \citenamefont
  {Dessau}}]{Kurleto23}%
  \BibitemOpen
  \bibfield  {author} {\bibinfo {author} {\bibfnamefont {R.}~\bibnamefont
  {Kurleto}}, \bibinfo {author} {\bibfnamefont {S.}~\bibnamefont {Lany}},
  \bibinfo {author} {\bibfnamefont {D.}~\bibnamefont {Pashov}}, \bibinfo
  {author} {\bibfnamefont {S.}~\bibnamefont {Acharya}}, \bibinfo {author}
  {\bibfnamefont {M.}~\bibnamefont {van Schilfgaarde}}, \ and\ \bibinfo
  {author} {\bibfnamefont {D.~S.}\ \bibnamefont {Dessau}},\ }\href@noop {}
  {\bibfield  {journal} {\bibinfo  {journal} {arXiv:2308.00698}\ } (\bibinfo
  {year} {2023})}\BibitemShut {NoStop}%
\bibitem [{\citenamefont {Cabezas-Escares}\ \emph {et~al.}(2023)\citenamefont
  {Cabezas-Escares}, \citenamefont {Barrera}, \citenamefont {Cardenas},\ and\
  \citenamefont {Munoz}}]{CabezasEscares23}%
  \BibitemOpen
  \bibfield  {author} {\bibinfo {author} {\bibfnamefont {J.}~\bibnamefont
  {Cabezas-Escares}}, \bibinfo {author} {\bibfnamefont {N.~F.}\ \bibnamefont
  {Barrera}}, \bibinfo {author} {\bibfnamefont {C.}~\bibnamefont {Cardenas}}, \
  and\ \bibinfo {author} {\bibfnamefont {F.}~\bibnamefont {Munoz}},\
  }\href@noop {} {\bibfield  {journal} {\bibinfo  {journal} {arXiv:2308.01135}\
  } (\bibinfo {year} {2023})}\BibitemShut {NoStop}%
\bibitem [{\citenamefont {Tavakol}\ and\ \citenamefont
  {Scaffidi}(2023)}]{Tavakol23}%
  \BibitemOpen
  \bibfield  {author} {\bibinfo {author} {\bibfnamefont {O.}~\bibnamefont
  {Tavakol}}\ and\ \bibinfo {author} {\bibfnamefont {T.}~\bibnamefont
  {Scaffidi}},\ }\href@noop {} {\bibfield  {journal} {\bibinfo  {journal}
  {arXiv:2308.01315}\ } (\bibinfo {year} {2023})}\BibitemShut {NoStop}%
\bibitem [{\citenamefont {Schnyder}\ \emph {et~al.}(2008)\citenamefont
  {Schnyder}, \citenamefont {Ryu}, \citenamefont {Furusaki},\ and\
  \citenamefont {Ludwig}}]{Schnyder08}%
  \BibitemOpen
  \bibfield  {author} {\bibinfo {author} {\bibfnamefont {A.~P.}\ \bibnamefont
  {Schnyder}}, \bibinfo {author} {\bibfnamefont {S.}~\bibnamefont {Ryu}},
  \bibinfo {author} {\bibfnamefont {A.}~\bibnamefont {Furusaki}}, \ and\
  \bibinfo {author} {\bibfnamefont {A.~W.~W.}\ \bibnamefont {Ludwig}},\ }\href
  {\doibase 10.1103/PhysRevB.78.195125} {\bibfield  {journal} {\bibinfo
  {journal} {Phys. Rev. B}\ }\textbf {\bibinfo {volume} {78}},\ \bibinfo
  {pages} {195125} (\bibinfo {year} {2008})}\BibitemShut {NoStop}%
\bibitem [{\citenamefont {Ryu}\ \emph {et~al.}(2010)\citenamefont {Ryu},
  \citenamefont {Schnyder}, \citenamefont {Furusaki},\ and\ \citenamefont
  {Ludwig}}]{Ryu10}%
  \BibitemOpen
  \bibfield  {author} {\bibinfo {author} {\bibfnamefont {S.}~\bibnamefont
  {Ryu}}, \bibinfo {author} {\bibfnamefont {A.~P.}\ \bibnamefont {Schnyder}},
  \bibinfo {author} {\bibfnamefont {A.}~\bibnamefont {Furusaki}}, \ and\
  \bibinfo {author} {\bibfnamefont {A.~W.~W.}\ \bibnamefont {Ludwig}},\ }\href
  {http://stacks.iop.org/1367-2630/12/i=6/a=065010} {\bibfield  {journal}
  {\bibinfo  {journal} {New J. Phys.}\ }\textbf {\bibinfo {volume} {12}},\
  \bibinfo {pages} {065010} (\bibinfo {year} {2010})}\BibitemShut {NoStop}%
\bibitem [{\citenamefont {Chiu}\ \emph {et~al.}(2016)\citenamefont {Chiu},
  \citenamefont {Teo}, \citenamefont {Schnyder},\ and\ \citenamefont
  {Ryu}}]{Chiu16}%
  \BibitemOpen
  \bibfield  {author} {\bibinfo {author} {\bibfnamefont {C.-K.}\ \bibnamefont
  {Chiu}}, \bibinfo {author} {\bibfnamefont {J.~C.~Y.}\ \bibnamefont {Teo}},
  \bibinfo {author} {\bibfnamefont {A.~P.}\ \bibnamefont {Schnyder}}, \ and\
  \bibinfo {author} {\bibfnamefont {S.}~\bibnamefont {Ryu}},\ }\href {\doibase
  10.1103/RevModPhys.88.035005} {\bibfield  {journal} {\bibinfo  {journal}
  {Rev. Mod. Phys.}\ }\textbf {\bibinfo {volume} {88}},\ \bibinfo {pages}
  {035005} (\bibinfo {year} {2016})}\BibitemShut {NoStop}%
\bibitem [{\citenamefont {Hasan}\ and\ \citenamefont {Kane}(2010)}]{Hasan10}%
  \BibitemOpen
  \bibfield  {author} {\bibinfo {author} {\bibfnamefont {M.~Z.}\ \bibnamefont
  {Hasan}}\ and\ \bibinfo {author} {\bibfnamefont {C.~L.}\ \bibnamefont
  {Kane}},\ }\href {\doibase 10.1103/RevModPhys.82.3045} {\bibfield  {journal}
  {\bibinfo  {journal} {Rev. Mod. Phys.}\ }\textbf {\bibinfo {volume} {82}},\
  \bibinfo {pages} {3045} (\bibinfo {year} {2010})}\BibitemShut {NoStop}%
\bibitem [{\citenamefont {Qi}\ and\ \citenamefont {Zhang}(2011)}]{Qi11}%
  \BibitemOpen
  \bibfield  {author} {\bibinfo {author} {\bibfnamefont {X.-L.}\ \bibnamefont
  {Qi}}\ and\ \bibinfo {author} {\bibfnamefont {S.-C.}\ \bibnamefont {Zhang}},\
  }\href {\doibase 10.1103/RevModPhys.83.1057} {\bibfield  {journal} {\bibinfo
  {journal} {Rev. Mod. Phys.}\ }\textbf {\bibinfo {volume} {83}},\ \bibinfo
  {pages} {1057} (\bibinfo {year} {2011})}\BibitemShut {NoStop}%
\bibitem [{\citenamefont {Chang}\ and\ \citenamefont {Niu}(1996)}]{Chang96}%
  \BibitemOpen
  \bibfield  {author} {\bibinfo {author} {\bibfnamefont {M.-C.}\ \bibnamefont
  {Chang}}\ and\ \bibinfo {author} {\bibfnamefont {Q.}~\bibnamefont {Niu}},\
  }\href {\doibase 10.1103/PhysRevB.53.7010} {\bibfield  {journal} {\bibinfo
  {journal} {Phys. Rev. B}\ }\textbf {\bibinfo {volume} {53}},\ \bibinfo
  {pages} {7010} (\bibinfo {year} {1996})}\BibitemShut {NoStop}%
\bibitem [{\citenamefont {Xiao}\ \emph {et~al.}(2010)\citenamefont {Xiao},
  \citenamefont {Chang},\ and\ \citenamefont {Niu}}]{Xiao10}%
  \BibitemOpen
  \bibfield  {author} {\bibinfo {author} {\bibfnamefont {D.}~\bibnamefont
  {Xiao}}, \bibinfo {author} {\bibfnamefont {M.-C.}\ \bibnamefont {Chang}}, \
  and\ \bibinfo {author} {\bibfnamefont {Q.}~\bibnamefont {Niu}},\ }\href
  {\doibase 10.1103/RevModPhys.82.1959} {\bibfield  {journal} {\bibinfo
  {journal} {Rev. Mod. Phys.}\ }\textbf {\bibinfo {volume} {82}},\ \bibinfo
  {pages} {1959} (\bibinfo {year} {2010})}\BibitemShut {NoStop}%
\bibitem [{\citenamefont {Cao}\ \emph {et~al.}(2018{\natexlab{a}})\citenamefont
  {Cao}, \citenamefont {Fatemi}, \citenamefont {Demir}, \citenamefont {Fang},
  \citenamefont {Tomarken}, \citenamefont {Luo}, \citenamefont
  {Sanchez-Yamagishi}, \citenamefont {Watanabe}, \citenamefont {Taniguchi},
  \citenamefont {Kaxiras}, \citenamefont {Ashoori},\ and\ \citenamefont
  {Jarillo-Herrero}}]{Cao18}%
  \BibitemOpen
  \bibfield  {author} {\bibinfo {author} {\bibfnamefont {Y.}~\bibnamefont
  {Cao}}, \bibinfo {author} {\bibfnamefont {V.}~\bibnamefont {Fatemi}},
  \bibinfo {author} {\bibfnamefont {A.}~\bibnamefont {Demir}}, \bibinfo
  {author} {\bibfnamefont {S.}~\bibnamefont {Fang}}, \bibinfo {author}
  {\bibfnamefont {S.~L.}\ \bibnamefont {Tomarken}}, \bibinfo {author}
  {\bibfnamefont {J.~Y.}\ \bibnamefont {Luo}}, \bibinfo {author} {\bibfnamefont
  {J.~D.}\ \bibnamefont {Sanchez-Yamagishi}}, \bibinfo {author} {\bibfnamefont
  {K.}~\bibnamefont {Watanabe}}, \bibinfo {author} {\bibfnamefont
  {T.}~\bibnamefont {Taniguchi}}, \bibinfo {author} {\bibfnamefont
  {E.}~\bibnamefont {Kaxiras}}, \bibinfo {author} {\bibfnamefont {R.~C.}\
  \bibnamefont {Ashoori}}, \ and\ \bibinfo {author} {\bibfnamefont
  {P.}~\bibnamefont {Jarillo-Herrero}},\ }\href {\doibase 10.1038/nature26154}
  {\bibfield  {journal} {\bibinfo  {journal} {Nature}\ }\textbf {\bibinfo
  {volume} {556}},\ \bibinfo {pages} {80} (\bibinfo {year}
  {2018}{\natexlab{a}})}\BibitemShut {NoStop}%
\bibitem [{\citenamefont {Cao}\ \emph {et~al.}(2018{\natexlab{b}})\citenamefont
  {Cao}, \citenamefont {Fatemi}, \citenamefont {Fang}, \citenamefont
  {Watanabe}, \citenamefont {Taniguchi}, \citenamefont {Kaxiras},\ and\
  \citenamefont {Jarillo-Herrero}}]{Cao18_2}%
  \BibitemOpen
  \bibfield  {author} {\bibinfo {author} {\bibfnamefont {Y.}~\bibnamefont
  {Cao}}, \bibinfo {author} {\bibfnamefont {V.}~\bibnamefont {Fatemi}},
  \bibinfo {author} {\bibfnamefont {S.}~\bibnamefont {Fang}}, \bibinfo {author}
  {\bibfnamefont {K.}~\bibnamefont {Watanabe}}, \bibinfo {author}
  {\bibfnamefont {T.}~\bibnamefont {Taniguchi}}, \bibinfo {author}
  {\bibfnamefont {E.}~\bibnamefont {Kaxiras}}, \ and\ \bibinfo {author}
  {\bibfnamefont {P.}~\bibnamefont {Jarillo-Herrero}},\ }\href {\doibase
  10.1038/nature26160} {\bibfield  {journal} {\bibinfo  {journal} {Nature}\
  }\textbf {\bibinfo {volume} {556}},\ \bibinfo {pages} {43} (\bibinfo {year}
  {2018}{\natexlab{b}})}\BibitemShut {NoStop}%
\bibitem [{\citenamefont {Peotta}\ and\ \citenamefont
  {T\"orm\"a}()}]{Peotta15}%
  \BibitemOpen
  \bibfield  {author} {\bibinfo {author} {\bibfnamefont {S.}~\bibnamefont
  {Peotta}}\ and\ \bibinfo {author} {\bibfnamefont {t.~j. y. m. d. v. n. p. i.
  d.~u.}\ \bibnamefont {T\"orm\"a}, \bibfnamefont {P\"aivi}},\ }\href@noop {}
  {\ }\BibitemShut {NoStop}%
\bibitem [{\citenamefont {Julku}\ \emph {et~al.}(2016)\citenamefont {Julku},
  \citenamefont {Peotta}, \citenamefont {Vanhala}, \citenamefont {Kim},\ and\
  \citenamefont {T\"orm\"a}}]{Julku16}%
  \BibitemOpen
  \bibfield  {author} {\bibinfo {author} {\bibfnamefont {A.}~\bibnamefont
  {Julku}}, \bibinfo {author} {\bibfnamefont {S.}~\bibnamefont {Peotta}},
  \bibinfo {author} {\bibfnamefont {T.~I.}\ \bibnamefont {Vanhala}}, \bibinfo
  {author} {\bibfnamefont {D.-H.}\ \bibnamefont {Kim}}, \ and\ \bibinfo
  {author} {\bibfnamefont {P.}~\bibnamefont {T\"orm\"a}},\ }\href {\doibase
  10.1103/PhysRevLett.117.045303} {\bibfield  {journal} {\bibinfo  {journal}
  {Phys. Rev. Lett.}\ }\textbf {\bibinfo {volume} {117}},\ \bibinfo {pages}
  {045303} (\bibinfo {year} {2016})}\BibitemShut {NoStop}%
\bibitem [{\citenamefont {Liang}\ \emph {et~al.}(2017)\citenamefont {Liang},
  \citenamefont {Vanhala}, \citenamefont {Peotta}, \citenamefont {Siro},
  \citenamefont {Harju},\ and\ \citenamefont {T\"orm\"a}}]{Liang17}%
  \BibitemOpen
  \bibfield  {author} {\bibinfo {author} {\bibfnamefont {L.}~\bibnamefont
  {Liang}}, \bibinfo {author} {\bibfnamefont {T.~I.}\ \bibnamefont {Vanhala}},
  \bibinfo {author} {\bibfnamefont {S.}~\bibnamefont {Peotta}}, \bibinfo
  {author} {\bibfnamefont {T.}~\bibnamefont {Siro}}, \bibinfo {author}
  {\bibfnamefont {A.}~\bibnamefont {Harju}}, \ and\ \bibinfo {author}
  {\bibfnamefont {P.}~\bibnamefont {T\"orm\"a}},\ }\href {\doibase
  10.1103/PhysRevB.95.024515} {\bibfield  {journal} {\bibinfo  {journal} {Phys.
  Rev. B}\ }\textbf {\bibinfo {volume} {95}},\ \bibinfo {pages} {024515}
  (\bibinfo {year} {2017})}\BibitemShut {NoStop}%
\bibitem [{\citenamefont {Herzog-Arbeitman}\ \emph {et~al.}(2022)\citenamefont
  {Herzog-Arbeitman}, \citenamefont {Peri}, \citenamefont {Schindler},
  \citenamefont {Huber},\ and\ \citenamefont {Bernevig}}]{HerzogArbeitman22}%
  \BibitemOpen
  \bibfield  {author} {\bibinfo {author} {\bibfnamefont {J.}~\bibnamefont
  {Herzog-Arbeitman}}, \bibinfo {author} {\bibfnamefont {V.}~\bibnamefont
  {Peri}}, \bibinfo {author} {\bibfnamefont {F.}~\bibnamefont {Schindler}},
  \bibinfo {author} {\bibfnamefont {S.~D.}\ \bibnamefont {Huber}}, \ and\
  \bibinfo {author} {\bibfnamefont {B.~A.}\ \bibnamefont {Bernevig}},\ }\href
  {\doibase 10.1103/PhysRevLett.128.087002} {\bibfield  {journal} {\bibinfo
  {journal} {Phys. Rev. Lett.}\ }\textbf {\bibinfo {volume} {128}},\ \bibinfo
  {pages} {087002} (\bibinfo {year} {2022})}\BibitemShut {NoStop}%
\bibitem [{\citenamefont {T\"{o}rm\"{a}}\ \emph {et~al.}(2022)\citenamefont
  {T\"{o}rm\"{a}}, \citenamefont {Peotta},\ and\ \citenamefont
  {Bernevig}}]{Torma22}%
  \BibitemOpen
  \bibfield  {author} {\bibinfo {author} {\bibfnamefont {P.}~\bibnamefont
  {T\"{o}rm\"{a}}}, \bibinfo {author} {\bibfnamefont {S.}~\bibnamefont
  {Peotta}}, \ and\ \bibinfo {author} {\bibfnamefont {B.~A.}\ \bibnamefont
  {Bernevig}},\ }\href {\doibase 10.1038/s42254-022-00466-y} {\bibfield
  {journal} {\bibinfo  {journal} {Nature Reviews Physics}\ }\textbf {\bibinfo
  {volume} {4}},\ \bibinfo {pages} {528} (\bibinfo {year} {2022})}\BibitemShut
  {NoStop}%
\bibitem [{\citenamefont {Sutherland}(1986)}]{Sutherland86}%
  \BibitemOpen
  \bibfield  {author} {\bibinfo {author} {\bibfnamefont {B.}~\bibnamefont
  {Sutherland}},\ }\href {\doibase 10.1103/PhysRevB.34.5208} {\bibfield
  {journal} {\bibinfo  {journal} {Phys. Rev. B}\ }\textbf {\bibinfo {volume}
  {34}},\ \bibinfo {pages} {5208} (\bibinfo {year} {1986})}\BibitemShut
  {NoStop}%
\bibitem [{\citenamefont {Lieb}(1989)}]{Lieb89}%
  \BibitemOpen
  \bibfield  {author} {\bibinfo {author} {\bibfnamefont {E.~H.}\ \bibnamefont
  {Lieb}},\ }\href {\doibase 10.1103/PhysRevLett.62.1201} {\bibfield  {journal}
  {\bibinfo  {journal} {Phys. Rev. Lett.}\ }\textbf {\bibinfo {volume} {62}},\
  \bibinfo {pages} {1201} (\bibinfo {year} {1989})}\BibitemShut {NoStop}%
\bibitem [{\citenamefont {de~Sousa}\ \emph {et~al.}(2022)\citenamefont
  {de~Sousa}, \citenamefont {Liu}, \citenamefont {Qu},\ and\ \citenamefont
  {Chen}}]{deSousa23_flat_band}%
  \BibitemOpen
  \bibfield  {author} {\bibinfo {author} {\bibfnamefont {M.~S.~M.}\
  \bibnamefont {de~Sousa}}, \bibinfo {author} {\bibfnamefont {F.}~\bibnamefont
  {Liu}}, \bibinfo {author} {\bibfnamefont {F.}~\bibnamefont {Qu}}, \ and\
  \bibinfo {author} {\bibfnamefont {W.}~\bibnamefont {Chen}},\ }\href {\doibase
  10.1103/PhysRevB.105.014511} {\bibfield  {journal} {\bibinfo  {journal}
  {Phys. Rev. B}\ }\textbf {\bibinfo {volume} {105}},\ \bibinfo {pages}
  {014511} (\bibinfo {year} {2022})}\BibitemShut {NoStop}%
\bibitem [{\citenamefont {Provost}\ and\ \citenamefont
  {Vallee}(1980)}]{Provost80}%
  \BibitemOpen
  \bibfield  {author} {\bibinfo {author} {\bibfnamefont {J.~P.}\ \bibnamefont
  {Provost}}\ and\ \bibinfo {author} {\bibfnamefont {G.}~\bibnamefont
  {Vallee}},\ }\href {https://projecteuclid.org:443/euclid.cmp/1103908308}
  {\bibfield  {journal} {\bibinfo  {journal} {Comm. Math. Phys.}\ }\textbf
  {\bibinfo {volume} {76}},\ \bibinfo {pages} {289} (\bibinfo {year}
  {1980})}\BibitemShut {NoStop}%
\bibitem [{\citenamefont {von Gersdorff}\ and\ \citenamefont
  {Chen}(2021)}]{vonGersdorff21_metric_curvature}%
  \BibitemOpen
  \bibfield  {author} {\bibinfo {author} {\bibfnamefont {G.}~\bibnamefont {von
  Gersdorff}}\ and\ \bibinfo {author} {\bibfnamefont {W.}~\bibnamefont
  {Chen}},\ }\href {\doibase 10.1103/PhysRevB.104.195133} {\bibfield  {journal}
  {\bibinfo  {journal} {Phys. Rev. B}\ }\textbf {\bibinfo {volume} {104}},\
  \bibinfo {pages} {195133} (\bibinfo {year} {2021})}\BibitemShut {NoStop}%
\bibitem [{\citenamefont {Chen}\ and\ \citenamefont {von
  Gersdorff}(2022)}]{Chen23_dressed_Berry_metric}%
  \BibitemOpen
  \bibfield  {author} {\bibinfo {author} {\bibfnamefont {W.}~\bibnamefont
  {Chen}}\ and\ \bibinfo {author} {\bibfnamefont {G.}~\bibnamefont {von
  Gersdorff}},\ }\href {\doibase 10.21468/SciPostPhysCore.5.3.040} {\bibfield
  {journal} {\bibinfo  {journal} {SciPost Phys. Core}\ }\textbf {\bibinfo
  {volume} {5}},\ \bibinfo {pages} {040} (\bibinfo {year} {2022})}\BibitemShut
  {NoStop}%
\bibitem [{\citenamefont {de~Sousa}\ \emph {et~al.}(2023)\citenamefont
  {de~Sousa}, \citenamefont {Cruz},\ and\ \citenamefont
  {Chen}}]{deSousa23_fidelity_marker}%
  \BibitemOpen
  \bibfield  {author} {\bibinfo {author} {\bibfnamefont {M.~S.~M.}\
  \bibnamefont {de~Sousa}}, \bibinfo {author} {\bibfnamefont {A.~L.}\
  \bibnamefont {Cruz}}, \ and\ \bibinfo {author} {\bibfnamefont
  {W.}~\bibnamefont {Chen}},\ }\href {\doibase 10.1103/PhysRevB.107.205133}
  {\bibfield  {journal} {\bibinfo  {journal} {Phys. Rev. B}\ }\textbf {\bibinfo
  {volume} {107}},\ \bibinfo {pages} {205133} (\bibinfo {year}
  {2023})}\BibitemShut {NoStop}%
\bibitem [{\citenamefont {Armitage}\ \emph {et~al.}(2018)\citenamefont
  {Armitage}, \citenamefont {Mele},\ and\ \citenamefont
  {Vishwanath}}]{Armitage18}%
  \BibitemOpen
  \bibfield  {author} {\bibinfo {author} {\bibfnamefont {N.~P.}\ \bibnamefont
  {Armitage}}, \bibinfo {author} {\bibfnamefont {E.~J.}\ \bibnamefont {Mele}},
  \ and\ \bibinfo {author} {\bibfnamefont {A.}~\bibnamefont {Vishwanath}},\
  }\href {\doibase 10.1103/RevModPhys.90.015001} {\bibfield  {journal}
  {\bibinfo  {journal} {Rev. Mod. Phys.}\ }\textbf {\bibinfo {volume} {90}},\
  \bibinfo {pages} {015001} (\bibinfo {year} {2018})}\BibitemShut {NoStop}%
\bibitem [{\citenamefont {Yan}\ and\ \citenamefont {Felser}(2017)}]{Yan17}%
  \BibitemOpen
  \bibfield  {author} {\bibinfo {author} {\bibfnamefont {B.}~\bibnamefont
  {Yan}}\ and\ \bibinfo {author} {\bibfnamefont {C.}~\bibnamefont {Felser}},\
  }\href {\doibase 10.1146/annurev-conmatphys-031016-025458} {\bibfield
  {journal} {\bibinfo  {journal} {Annu. Rev. Condens. Matter Phys.}\ }\textbf
  {\bibinfo {volume} {8}},\ \bibinfo {pages} {337} (\bibinfo {year}
  {2017})}\BibitemShut {NoStop}%
\bibitem [{\citenamefont {Lv}\ \emph {et~al.}(2021)\citenamefont {Lv},
  \citenamefont {Qian},\ and\ \citenamefont {Ding}}]{Lv21}%
  \BibitemOpen
  \bibfield  {author} {\bibinfo {author} {\bibfnamefont {B.~Q.}\ \bibnamefont
  {Lv}}, \bibinfo {author} {\bibfnamefont {T.}~\bibnamefont {Qian}}, \ and\
  \bibinfo {author} {\bibfnamefont {H.}~\bibnamefont {Ding}},\ }\href {\doibase
  10.1103/RevModPhys.93.025002} {\bibfield  {journal} {\bibinfo  {journal}
  {Rev. Mod. Phys.}\ }\textbf {\bibinfo {volume} {93}},\ \bibinfo {pages}
  {025002} (\bibinfo {year} {2021})}\BibitemShut {NoStop}%
\bibitem [{\citenamefont {Feng}\ \emph {et~al.}(2012)\citenamefont {Feng},
  \citenamefont {Yao}, \citenamefont {Zhu}, \citenamefont {Zhou}, \citenamefont
  {Yao},\ and\ \citenamefont {Xiao}}]{Feng12}%
  \BibitemOpen
  \bibfield  {author} {\bibinfo {author} {\bibfnamefont {W.}~\bibnamefont
  {Feng}}, \bibinfo {author} {\bibfnamefont {Y.}~\bibnamefont {Yao}}, \bibinfo
  {author} {\bibfnamefont {W.}~\bibnamefont {Zhu}}, \bibinfo {author}
  {\bibfnamefont {J.}~\bibnamefont {Zhou}}, \bibinfo {author} {\bibfnamefont
  {W.}~\bibnamefont {Yao}}, \ and\ \bibinfo {author} {\bibfnamefont
  {D.}~\bibnamefont {Xiao}},\ }\href {\doibase 10.1103/PhysRevB.86.165108}
  {\bibfield  {journal} {\bibinfo  {journal} {Phys. Rev. B}\ }\textbf {\bibinfo
  {volume} {86}},\ \bibinfo {pages} {165108} (\bibinfo {year}
  {2012})}\BibitemShut {NoStop}%
\bibitem [{\citenamefont {Liu}\ \emph {et~al.}(2013)\citenamefont {Liu},
  \citenamefont {Shan}, \citenamefont {Yao}, \citenamefont {Yao},\ and\
  \citenamefont {Xiao}}]{Liu13}%
  \BibitemOpen
  \bibfield  {author} {\bibinfo {author} {\bibfnamefont {G.-B.}\ \bibnamefont
  {Liu}}, \bibinfo {author} {\bibfnamefont {W.-Y.}\ \bibnamefont {Shan}},
  \bibinfo {author} {\bibfnamefont {Y.}~\bibnamefont {Yao}}, \bibinfo {author}
  {\bibfnamefont {W.}~\bibnamefont {Yao}}, \ and\ \bibinfo {author}
  {\bibfnamefont {D.}~\bibnamefont {Xiao}},\ }\href {\doibase
  10.1103/PhysRevB.88.085433} {\bibfield  {journal} {\bibinfo  {journal} {Phys.
  Rev. B}\ }\textbf {\bibinfo {volume} {88}},\ \bibinfo {pages} {085433}
  (\bibinfo {year} {2013})}\BibitemShut {NoStop}%
\bibitem [{\citenamefont {Molignini}\ \emph {et~al.}(2022)\citenamefont
  {Molignini}, \citenamefont {Lapierre}, \citenamefont {Chitra},\ and\
  \citenamefont {Chen}}]{Molignini22_Chern_marker}%
  \BibitemOpen
  \bibfield  {author} {\bibinfo {author} {\bibfnamefont {P.}~\bibnamefont
  {Molignini}}, \bibinfo {author} {\bibfnamefont {B.}~\bibnamefont {Lapierre}},
  \bibinfo {author} {\bibfnamefont {R.}~\bibnamefont {Chitra}}, \ and\ \bibinfo
  {author} {\bibfnamefont {W.}~\bibnamefont {Chen}},\ }\href@noop {} {\bibfield
   {journal} {\bibinfo  {journal} {arXiv:2207.00016}\ } (\bibinfo {year}
  {2022})}\BibitemShut {NoStop}%
\end{thebibliography}%

\end{document}